\documentclass[aps,pra,showpacs,floatfix,superscriptaddress,reprint]{revtex4-1}
\usepackage[T1]{fontenc}
\usepackage{graphicx}
\usepackage{bm}
\usepackage{multirow}
\usepackage{braket}
\usepackage{color}
\usepackage{txfonts}
\usepackage{times}
\usepackage{hyperref}

\begin{document}
\title{Scattering lengths in isotopologues of the RbYb system}

\author{Mateusz Borkowski}
\author{Piotr S. \.Zuchowski}
\email{pzuch@fizyka.umk.pl}%
\author{Roman Ciury\l o}
\email{rciurylo@fizyka.umk.pl}
\affiliation{Institute of Physics, Faculty of Physics, Astronomy and Informatics, Nicolaus Copernicus University, Grudziadzka 5, 87-100 Torun, Poland}
\author{Paul S. Julienne}
\affiliation{Joint Quantum Institute, National Institute of Standards and Technology and the University of
Maryland, 100 Bureau Drive, Stop 8423, Gaithersburg, Maryland 20899-8423, USA.}
\author{Dariusz K\k{e}dziera}
\email{teodar@chem.umk.pl}
\affiliation{Department of Chemistry, Nicolaus Copernicus University, 7 Gagarin Street, 87-100 Torun, Poland}
\author{\L{}ukasz Mentel}
\affiliation{Section of Theoretical Chemistry, VU University, Amsterdam, The Netherlands}
\author{Pawe\l{} Tecmer}
\affiliation{ETH Zurich, Laboratory of Physical Chemistry, Wolfgang-Pauli-Str. 10, CH-8093 Z\"urich, Switzerland}
\author{Frank M\"unchow}
\author{Cristian Bruni}
\author{Axel G\"orlitz}
\affiliation{Institut f\"ur Experimentalphysik, Heinrich-Heine-Universit\"at
D\"usseldorf, Universit\"atstra{\ss}e 1, 40225 D\"usseldorf, Germany}

\date{\today}

\begin{abstract}
We model the binding energies of rovibrational levels of the RbYb molecule using experimental data from two-color photoassociation spectroscopy in mixtures of ultracold $^{87}$Rb with various Yb isotopes.  The model uses a theoretical potential based on state-of-the-art \emph{ab initio} potentials, further improved by least-squares fitting to the experimental data. We have fixed the number of bound states supported by the potential curve, so that the model is mass scaled, that is, it accurately describes the bound state energies for all measured isotopic combinations. Such a model enables an accurate prediction of the s-wave scattering lengths of all isotopic combinations of the RbYb system. The reduced mass range is broad enough to cover the full scattering lengths range from $-\infty$ to $+\infty$. For example, the $^{87}$Rb$^{174}$Yb system is characterized by a large positive scattering length of $+880(120)$~a.u., while $^{87}$Rb$^{173}$Yb has $a=-626(88)$~a.u.. On the other hand $^{87}$Rb$^{170}$Yb has a very small scattering length of $-14.5(1.8)$~a.u. confirmed by the pair's extremely low thermalization rate. For isotopic combinations including $^{85}$Rb the variation of the interspecies scattering lengths is much smoother ranging from $+39.0(1.6)$ a.u. for  $^{85}$Rb$^{176}$Yb to $+230(10)$ a.u. in the case of $^{85}$Rb$^{168}$Yb. Hyperfine corrections to these scattering lengths are also given. We further complement the fitted potential with interaction parameters calculated from alternative methods. The recommended value of the van der Waals coefficient is $C_6$=2837(13)~a.u. and is in agreement and more precise than the current state-of-the-art theoretical predictions (S. G. Porsev, M. S. Safronova, A. Derevianko, and C.W. Clark, arXiv:1307.2654  (2013)).
\end{abstract}

\maketitle


\section{Introduction}

Creation of ground-state polar molecules in the sub-microkelvin regime \cite{Ni:KRb:2008} is one of the most 
important achievements in atomic, molecular and optical physics in recent years. It is expected that further 
development of production techniques for ultracold molecular samples will find many extremely 
exciting applications, for example, in quantum information theory \cite{DeMille:2002}, quantum simulations of many-body physics \cite{Baranov:2012},
and high-precision measurements \cite{Zelevinsky:2008:test,Bethlem:2009,Kajita:2011}.

Production of polar molecules with non-zero electronic spin could open new exciting directions of research. 
For example, molecules that contain atoms with very large atomic number are considered as candidates
for investigations of limits of the electric dipole moment of the electron  \cite{meyer:2008,hudson:2002}.
Paramagnetic polar molecules have also been proposed as candidates for creating topologically ordered states 
and a new class of quantum simulators \cite{Micheli:2006}. Also, paramagnetic polar molecules could open new pathways in studies of chemical reactions at ultralow temperatures.  In contrast to alkali-metal dimers in their absolute ground states, paramagnetic and  polar molecular states are magnetically tunable, which  with electric field control might open new possibilites of manipulating chemical reactions by combined fields \cite{Ospelkaus:react:2010,Krems:05rev,Tscherbul:2006}.

There are ongoing efforts to produce ultracold polar molecules with nonzero spin directly, such as, for example, OH, NH, CaH or SrF by using Stark deceleration \cite{Hoekstra:2007,vandeMeerakker:2003,Bochinski:2003},
buffer-gas cooling \cite{Campbell:2007,Weinstein:CaH:1998} or, more recently, laser cooling \cite{Shuman:2010}. Despite these efforts direct cooling of molecules into the microkelvin regime is still not achieved. 

An alternative approach to obtain ultracold paramagnetic and polar molecules is to produce them from translationally ultracold atoms with differing numbers of electrons. 
The natural candidates for such molecules are pairs which combine an alkaline earth, or similar atom, such as Ca, Sr, Mg, Yb or Hg, and an alkali-metal atom. For both of these classes
of atoms techniques for magneto-optical and optical trapping, as well as internal state manipulation are well developed, and Bose-Einstein condensates have been obtained for most of these species.    

Currently, several systems of mixed alkaline-earth and alkali-metal atoms are intensely being studied by several research groups \cite{Nemitz:2009,Munchow:2011,Ivanov:2011,Hansen:2011,Hara:2011,Okano:2010} and very recently first quantum degenerate mixtures in such systems have been produced using Sr and Rb \cite{Pasquiou:2013}.
In the present work our investigations are focused on the Rb-Yb system. To date Rb and Yb atoms have been cotrapped in an optical dipole trap or a hybrid trap which combines a magnetic trap for Rb with an optical trap for Yb. The intraspecies scattering lengths for both Rb and Yb are well known --- $^{87}$Rb has a scattering length $a \approx 100$~a.u., while for $^{85}$Rb it is resonant \cite{Boesten:1997, Tsai:1997}; on the other hand, ytterbium isotopes span a full cycle of scattering lengths \cite{Kitagawa:2008}. It was possible to study the  process of mutual thermalization of the two atomic species, thus allowing estimates of the scattering lengths to be made for several Rb-Yb isotopic mixtures \cite{Baumer:2011,Tassy:2010}. In addition, the single-photon photoassociation 
spectrum near the Rb($5p$)+Yb($5s$) asymptote has been investigated \cite{Nemitz:2009}. More recently two-color photoassociation spectroscopy has been performed for the  $^{87}$Rb$^{176}$Yb molecular system  \cite{Munchow:2011} paving the way toward accurate determination of interaction potentials and scattering lengths of Rb-Yb mixtures. 

Another system being actively explored by experimentalists is the Li-Yb mixture.
The group at the  University of Washington \cite{Ivanov:2011,Hansen:2011}
sympathetically cooled Li atoms by collisions with ultracold Yb atoms below the Fermi temperature and estimated  the magnitude of the scattering length. The same system was investigated by the group of Kyoto University \cite{Hara:2011,Okano:2010} and the
magnitude of the scattering length found for $^6$Li$^{174}$Yb  confirmed the findings of Ivanov {\em et al.} \cite{Ivanov:2011}.
This system, contrary to Rb-Yb, is limited in its range of available scattering lengths because of the much smaller range of variation in reduced mass due to a much larger mass imbalance.  
 
The finding that also in mixtures of alkaline-earth and alkali atoms a mechanism for magnetic tunability of scattering lengths via Feshbach resonances exists \cite{Zuchowski:2010:RbSr,Brue:2012} has further strengthened the interest in such ultracold mixtures. Recently, a theoretical investigation of Feshbach resonances in ultracold mixtures of Yb and various alkaline species has been performed \cite{Brue:2013} which is in part based on the experimental data \cite{MunchowPhD:2012} which are analyzed and modelled within this manuscript. However, Feshbach resonances in such systems are predicted to be very narrow
and production of Feshbach molecules in such systems might be experimentally challenging. On the other hand, the recently reported formation of Sr$_2$ in electronic ground state \cite{Stellmer:2012} by stimulated Raman adiabatic passage (STIRAP) from  atom pairs trapped on sites of an optical lattice, demonstrated  a possibility to eliminate the necessity of using magnetic Feshbach resonances as a first step in production of ultracold molecules and it might be feasible to apply a similar scheme to RbYb. Both approaches require the precise knowledge of the molecular structure \cite{Skomorowski:2012}
in the ground and excited electronic states provided by ordinary molecular spectroscopy, as well as by one- and two-color photoassociation spectroscopy \cite{Zelevinsky:2006}. 

In this paper we present experimentally determined binding energies of the rovibrational levels of the $^2\Sigma_{1/2}$ electronic ground state of RbYb close to the dissociation limit for several isotopic combinations. These experimental data, obtained using two-photon photoassociation spectroscopy, are combined with state-of-the-art {\em ab initio} calculations to model the interaction potential for the Rb-Yb system with an accuracy high enough to predict scattering lengths for all isotopic combinations.  We explore mass-scaling of the phase variation of the scattering wavefunction in the ground electronic state for different isotopic mixtures of the Rb-Yb system. We propose a model potential of a form similar to the previous study of the RbSr dimer \cite{Zuchowski:2010:RbSr}, which employs an {\em ab initio} representation at short range smoothly connected \cite{Janssen:2009} to the long-range analytical form, which includes $C_6$ and $C_8$ van der Waals coefficients. However, we use the recorded spectra of the bound-state energies for several isotopic combinations of the Rb-Yb system and rotational quantum numbers $R=0,1$ to fix the total number of bound states as well as the position of the last bound states near the threshold. We provide a theoretical analysis of $C_6$ and $C_8$ coefficients based on the experimental data and compare them to the most recent theoretical calculations \cite{Safronova:2012, Porsev:2013}. Then we use the potential to interpret previously reported interspecies thermalization in Rb-Yb mixtures \cite{Baumer:2011, Tassy:2010}. The accurate potential reported will be very useful for future theoretical and experimental studies of magnetic and optical Feshbach resonances in Rb-Yb mixtures.

\section{Experiment}

\begin{figure}
\includegraphics[width=0.45\textwidth, clip]{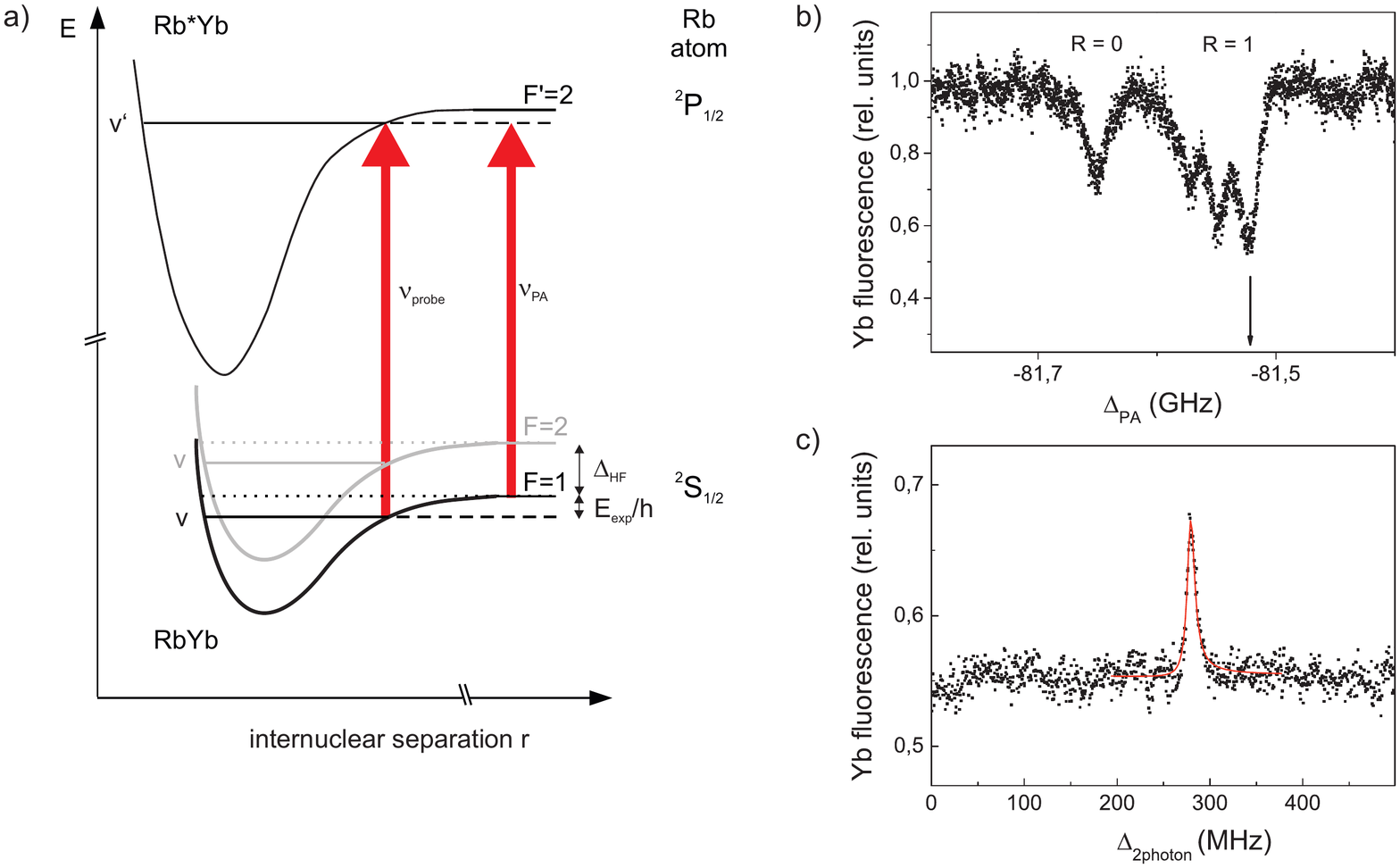}
\caption{a)\,Principle of two-photon photoassociation spectroscopy. b)\,One-photon photoassociation spectrum of a vibrational level of the $^2\Pi_{1/2}$ state excited state at $\Delta_{\rm PA} = -82$ GHz. c)\,Two-photon photoassociation spectrum of a vibrational level of the electronic ground state of $^{87}$Rb$^{176}$Yb with an experimentally measured two-photon transition frequency of $\Delta_{\rm 2photon} \approx 283$ MHz. For this mesurement the PA laser is set to the frequency indicated by the arrow in b). Thus, the probe laser connects vibrational levels with a rotational quantum number R=1.}
\label{fig:experiment}
\end{figure}

To determine the binding energies of rovibrational levels in the ground state of RbYb we have used two-photon photoassociation spectroscopy. The experimental results which are listed in table \ref{tab:data} represent an extension of previous measurements \cite{Munchow:2011} which only included binding energies of weakly bound rovibrational levels of the isotopologue $^{87}$Rb$^{176}$Yb with  nuclear rotation quantum number $R = 1$. Here also binding energies for rovibrational levels with $R = 0$ as well as for other isotopologues of RbYb are included.

The experimental determination of the binding energies follows the procedure which was already used in \cite{Munchow:2011} and is described in detail in \cite{MunchowPhD:2012}.  Two-photon photoassociation spectroscopy is performed in a double-species magneto-optical trap (MOT) employing a trap-loss technique. In steady state $\approx 10^8$ $^{87}$Rb atoms and $\approx 10^5$ Yb atoms (either $^{176}$Yb, $^{174}$Yb, $^{172}$Yb or $^{170}$Yb) are trapped in the continuously loaded double-species MOT at a temperature of a few hundred $\mu K$. Since the Rb MOT is operated in a so-called dark-spot configuration \cite{Anderson:1994}, more than 95$\%$ of the $^{87}$Rb atoms in the MOT are in the $F=1$ hyperfine level of the $^2S_{1/2}$ ground state.

To study heteronuclear photoassociation, the double-species MOT is first exposed to a tunable laser (PA-laser) with a wavelength close to the rubidium $^2S_{1/2} \rightarrow ^2P_{1/2}$ transition wavelength of $795$\,nm (PA-laser in Fig. \ref{fig:experiment}\, a)). If the frequency of the PA-laser matches the resonance condition for a transition from an unbound atom pair to an  excited Rb*Yb molecule in a specific weakly-bound rovibrational level of the electronically excited $^2\Pi_{1/2}$ state, the same number of Rb and Yb atoms is removed from the magnetooptical trap \cite{Nemitz:2009}. The corresponding reduction of the Yb steady-state atom number is detected as a reduction of the fluorescence of the Yb MOT operating at 556\,nm. In Fig. \ref{fig:experiment}\,b), the fluorescence signal of the Yb MOT corresponding to a transition to the $\Delta v'=-11$  vibrational level \footnote{The quantum number $\Delta v'$ is defined as $\Delta v' = v_{\rm max} - v$, where  $v$ is the vibrational quantum number and $v_{\rm max}$ is the vibrational quantum number of the most weakly bound vibrational level.} is depicted. The spectrum shows a resolved rotational structure as well as a splitting of the rotational components. No effect on the Rb MOT due to the formation of excited state Rb*Yb is observable since the Rb atom number in the MOT exceeds the Yb atom number by three orders of magnitude. 

To measure the ground state binding energy by two-photon photoassociation, a specific rovibrational level of the excited state is used as an intermediate state by fixing the PA-laser frequency to the resonance frequency for the corresponding one-photon photoassociation transition. An additional tunable laser (probe laser in Fig. \ref{fig:experiment}\, a)) is then applied to the double-species MOT. If the frequency difference $\Delta_{2photon}$ between the PA-laser and the probe laser matches the frequency (energy) difference between unbound atoms and a rovibrational level of the $^2\Sigma_{1/2}$ ground state of RbYb the production of excited state molecules is suppressed  \cite{Abraham:1995} via an Autler-Townes mechanism. This results in reduced atom loss and therefore the fluorescence of the Yb MOT increases. The positions $\Delta_{\rm PA}$ of the intermediate  levels of the $^2\Pi_{1/2}$ excited state used for each of the measured ground state energy levels are listed in Table~\ref{tab:data}. The values and error bars given in table I are obtained from a reanalysis and of the data presented in \cite{Munchow:2011} and \cite{MunchowPhD:2012} including additional data sets.

Each rovibrational level in the $^2\Sigma_{1/2}$ ground state of RbYb splits into two hyperfine levels which can be derived from the ground state hyperfine levels of atomic $^{87}$Rb. Due to the absence of any angular momentum in the $^1S_0$ ground state of the bosonic Yb isotopes that have been used in the present study, the hyperfine splitting of the dissociation limit in the ground state of the RbYb molecule may be assumed to be identical to the hyperfine splitting of atomic $^{87}$Rb of $6.835$\,GHz \cite{Bize:1999}. Thus, the experimentally determined detuning from the atomic line $\Delta_{\rm exp}$ corresponding to an observed two-photon-transition is given by 
\begin{equation}
 \Delta_{\rm exp}\,=\,\Delta_{\rm 2photon}\,+\,\Delta_{\rm HF} \,.
\end{equation}
The values for $\Delta_{\rm HF}$ are $\Delta_{\rm HF}(F=1) = 0$\,GHz and $\Delta_{\rm HF}(F=2) = - 6.835$\,GHz depending on whether the rovibrational level in the molecular ground state corresponds to the $^{87}$Rb atom being in the $F=1$ or the $F=2$ hyperfine level of the electronic ground state. The assignment of the observed two-photon resonances to the molecular hyperfine state is made in such a way that the binding energies agree with a simplified van der Waals model potential in accordance with the assumption that the hyperfine splitting of the weakly bound vibrational levels in the RbYb molecule is the same as in the $^{87}$Rb atom. In further analysis, however, we take the $r$-dependence of the hyperfine splitting into consideration.
 
Furthermore, the rotational quantum number of the rovibrational ground state level can be selected in two-photon photoassociation, by addressing a specific rotational level in the intermediate excited state. The reason is that the very weakly bound vibrational levels which are examined here are only coupled significantly by the probe laser if the rotational quantum number of the ground and the excited state are equal.     

The line centers of two-photon photoassociation spectra as depicted in Fig. \ref{fig:experiment}\,c) are found by fitting the spectra with an appropriate lineshape function \cite{MunchowPhD:2012, Bohn:1996}.  Temperature effects on the lineshape are assumed to be small compared to other experimental uncertainties, therefore they are neglected for the determination of the position of the line centers. Temperature shifts are generally on the order of the natural linewidth, or a few MHz here, depending on the partial wave; see Fig. 5 of Ref.~\cite{Jones:2000}. The temperature shift of the line center will be taken into account in the data analysis (see Section \ref{sec:results}), as we use the sample temperature $T$ as a fitting parameter. 

In order to determine the absolute value of $\Delta_{\rm 2photon}$, two different methods were used. For $\Delta_{\rm 2photon} < 2$ GHz, a beat signal of the probe and the PA-laser was recorded in most cases using a fast photodiode and a spectrum analyzer. For this method the resulting standard error of $\Delta_{\rm 2photon}$ and correspondingly  the line position $\Delta_{\rm exp}$ is estimated to be on the order of 10~MHz. For $\Delta_{\rm 2photon} > 2$ GHz the frequency difference between the two lasers was determined by measuring the wavelengths of the two lasers independently using a home-built wavemeter, that is based on a Michelson interferometer. This method is significantly less accurate and correspondingly the standard error for the determination of $\Delta_{\rm 2photon}$ is estimated to be a few hundred MHz. The experimental errors given in table \ref{tab:data} are estimated for each individual data set independently, taking into account the data quality and the specific experimental conditions..    

\begin{table*}[htbp]
  \centering
  \caption{Two color photoassociation line positions for RbYb measured in the experiment. The `Yb' column specifies the ytterbium isotope for the corresponding line; the other atom is $^{87}$Rb in all cases. $\Delta_{\rm PA}$ is the position of the intermediate excited state measured from the rubidium atomic D1 transition. The vibrational, rotational, and total angular momentum quantum numbers are labelled $v'$\footnote{In this paper the vibrational quantum number $v$ is counted from the dissociation limit with $v = 1$ being the most weakly bound state. In contrast, in previous RbYb papers \cite{Munchow:2011, MunchowPhD:2012} $\Delta \nu = -v +1$ was used instead to denote the vibrational state. }, $R$ and $F$, respectively. The experimentally measured line position $\Delta$ is given in the column labelled `Exp.' and its experimental uncertainty is in the column `Error'. The theoretical line position calculated from our best fit potential and sample temperature $T$ is given in the column `Theory'. The difference between the fitted and experimental line positions are labelled `Diff.'. Finally, in the last column, we give the theoretical bound state energies $E^{\rm th}$.}
\begin{ruledtabular}
    \begin{tabular}{rrrrrrrrrr|rrrrrrrrrr}
    Yb & $\Delta_{\rm PA} $ & $v$    & $R$     & $F$     & \multicolumn{4}{c}{$\Delta$ (MHz)} & $E_{\rm th}/h$ & Yb & $\Delta_{\rm PA}$& $v$    & $R$     & $F$      & \multicolumn{4}{c}{$\Delta$ (MHz)} & $E_{\rm th}/h$\\
          &   (GHz)     &       &       &       & Exp. & Error & Theory & Diff. &   (MHz)     &       &   (GHz)     &       &       &       & Exp. & Error & Theory & Diff. &   (MHz)\\
\hline
    170 & -24 & 1    & 0     & 1     & -113.3 & 15.0  & -106.0 & -7.3  & -102.9 & 176 & -147 & 4    & 0     & 2     & -5249.7 & 20.0  & -5250.1 & 0.4   & -5247.0 \\
    170 & -24 & 1    & 1     & 1     & -97.4 & 15.0  & -94.9 & -2.5  & -85.6 & 176 & -82 & 4    & 1     & 1     & -5203.7 & 143.7 & -5197.4 & -6.3  & -5188.2 \\
    170 & -24 & 2    & 0     & 1     & -1028.8 & 15.8  & -1021.6 & -7.2  & -1018.6 & 176 & -147 & 4    & 1     & 1     & -5190.9 & 100.0 & -5197.4 & 6.6   & -5188.2 \\
    170 & -24 & 2    & 1     & 1     & -990.8 & 9.5   & -990.9 & 0.1   & -981.7 & 176 & -82 & 4    & 1     & 2     & -5196.3 & 14.9  & -5194.3 & -2.0  & -5185.0 \\
    172 & -29 & 1    & 0     & 1     & -166.0 & 10.0  & -158.2 & -7.8  & -155.1 & 176 & -147 & 4    & 1     & 2     & -5190.2 & 19.8  & -5194.3 & 4.1   & -5185.0 \\
    172 & -64 & 1    & 0     & 1     & -153.8 & 20.0  & -158.2 & 4.4   & -155.1 & 176 & -147 & 5    & 0     & 1     & -11760.3 & 220.5 & -11754.0 & -6.3  & -11750.9 \\
    172 & -29 & 1    & 1     & 1     & -146.1 & 10.0  & -144.6 & -1.5  & -135.3 & 176 & -147 & 5    & 0     & 2     & -11765.5 & 150.0 & -11748.6 & -16.9 & -11745.5 \\
    172 & -29 & 2    & 0     & 1     & -1237.1 & 15.0  & -1239.7 & 2.6   & -1236.6 & 176 & -82 & 5    & 1     & 1     & -11697.3 & 257.2 & -11679.9 & -17.4 & -11670.6 \\
    172 & -29 & 2    & 1     & 1     & -1210.7 & 15.0  & -1206.7 & -4.0  & -1197.4 & 176 & -147 & 5    & 1     & 1     & -11673.7 & 212.1 & -11679.9 & 6.2   & -11670.6 \\
    174 & -34 & 1    & 0     & 1     & -227.9 & 10.0  & -223.8 & -4.1  & -220.8 & 176 & -147 & 5    & 1     & 2     & -11642.2 & 100.0 & -11674.5 & 32.3  & -11665.2 \\
    174 & -34 & 1    & 1     & 1     & -206.7 & 9.8   & -207.8 & 1.1   & -198.6 & 176 & -82 & 6    & 0     & 1     & -22174.8 & 116.7 & -22184.8 & 10.0  & -22181.7 \\
    174 & -34 & 2    & 0     & 1     & -1477.6 & 10.0  & -1481.4 & 3.8   & -1478.3 & 176 & -82 & 6    & 0     & 2     & -22207.4 & 94.5  & -22176.5 & -30.9 & -22173.4 \\
    174 & -34 & 2    & 1     & 1     & -1443.7 & 18.6  & -1446.2 & 2.5   & -1437.0 & 176 & -82 & 6    & 1     & 1     & -22082.9 & 143.7 & -22092.6 & 9.7   & -22083.4 \\
    176 & -82 & 2    & 0     & 1     & -304.2 & 7.7   & -303.8 & -0.4  & -300.7 & 176 & -82 & 6    & 1     & 2     & -22132.0 & 73.1  & -22084.4 & -47.6 & -22075.1 \\
    176 & -147 & 2    & 0     & 1     & -304.9 & 7.1   & -303.8 & -1.0  & -300.7 & 176 & -82 & 7    & 0     & 1     & -37487.4 & 120.0 & -37479.1 & -8.3  & -37476.0 \\
    176 & -82 & 2    & 1     & 1     & -283.4 & 7.2   & -285.5 & 2.1   & -276.3 & 176 & -147 & 7    & 0     & 1     & -37292.7 & 212.1 & -37479.1 & 186.5 & -37476.0 \\
    176 & -147 & 2    & 1     & 1     & -291.6 & 14.9  & -285.5 & -6.1  & -276.3 & 176 & -82 & 7    & 0     & 2     & -37446.2 & 155.7 & -37467.3 & 21.1  & -37464.3 \\
    176 & -82 & 2    & 1     & 2     & -304.2 & 134.2 & -285.1 & -19.1 & -275.8 & 176 & -147 & 7    & 0     & 2     & -37431.7 & 220.5 & -37467.3 & 35.6  & -37464.3 \\
    176 & -82 & 3    & 0     & 1     & -1748.6 & 10.0  & -1747.0 & -1.6  & -1743.9 & 176 & -82 & 7    & 1     & 1     & -37395.9 & 120.0 & -37369.1 & -26.8 & -37359.8 \\
    176 & -147 & 3    & 0     & 1     & -1748.5 & 7.3   & -1747.0 & -1.6  & -1743.9 & 176 & -147 & 7    & 1     & 1     & -37288.0 & 212.1 & -37369.1 & 81.1  & -37359.8 \\
    176 & -82 & 3    & 1     & 1     & -1700.2 & 10.0  & -1709.7 & 9.5   & -1700.5 & 176 & -82 & 7    & 1     & 2     & -37381.0 & 143.7 & -37357.3 & -23.7 & -37348.1 \\
    176 & -147 & 3    & 1     & 1     & -1712.1 & 14.9  & -1709.7 & -2.4  & -1700.5 & 176 & -147 & 7    & 1     & 2     & -37315.1 & 212.1 & -37357.3 & 42.2  & -37348.1 \\
    176 & -82 & 3    & 1     & 2     & -1748.5 & 134.2 & -1708.2 & -40.3 & -1699.0 & 176 & -147 & 8    & 0     & 1     & -58557.4 & 84.9  & -58562.8 & 5.5   & -58559.7 \\
    176 & -147 & 3    & 1     & 2     & -1754.1 & 200.0 & -1708.2 & -45.9 & -1699.0 & 176 & -147 & 8    & 0     & 2     & -58430.3 & 212.1 & -58547.0 & 116.7 & -58543.9 \\
    176 & -82 & 4    & 0     & 1     & -5235.0 & 155.7 & -5253.2 & 18.2  & -5250.2 & 176 & -147 & 8    & 1     & 1     & -58452.4 & 93.7  & -58435.1 & -17.2 & -58425.9 \\
    176 & -147 & 4    & 0     & 1     & -5277.5 & 150.0 & -5253.2 & -24.3 & -5250.2 & 176 & -147 & 8    & 1     & 2     & -58455.4 & 134.2 & -58419.3 & -36.2 & -58410.0 \\
    \end{tabular}%
\end{ruledtabular}
  \label{tab:data}%
\end{table*}%

\section{Mass scaling}

The aim of this work is to obtain accurate values of the interspecies scattering lengths of all Rb-Yb isotopic combinations based on the knowledge of the energy levels of only four such systems. To this end it is necessary to extrapolate the collisional properties to other isotopologues via mass-scaling, that is, the use of the same interaction potential $V(r)$ for all isotopic combinations. In this section we will show the theoretical foundation for such a procedure. A similar approach has been used in previous determinations of scattering lengths in e.g. Sr \cite{deEscobar:2008}, Yb \cite{Kitagawa:2008} and Rb atoms \cite{Tsai:1997}. 

In the limit of zero collision energy the scattering length for the 
interaction potential asymptotically dominated by a van der Waals $-C_6r^{-6}$
term can be  well approximated by ~\cite{Gribakin:1993} 
\begin{equation}
a=\bar{a}\left[1- \tan \left ( \Phi -\frac{\pi}{8}\right) \right], \label{eq:gribakin}
\end{equation}
where $\bar{a}= 2^{-\frac{3}{2}}\frac{\Gamma(3/4)}{\Gamma(5/4)}(2\mu C_6/ \hbar^2)^\frac{1}{4} $
is the characteristic length scale associated with the interaction strength of the colliding atoms,
$C_6$ is the van der Waals coefficient, $\mu = \left( m_1^{-1}+m_2^{-1} \right)^{-1}$ is the reduced mass and $\Phi$ is a zero energy WKB phase integral,
written in terms of the Born-Oppenheimer (BO) interaction potential $V(r)$:
\begin{equation}
\Phi=\frac{\sqrt{2\mu}}{\hbar} \int^{\infty}_{r_{\rm in}} \sqrt{ -V(r) } {\rm d} r.
\label{Phi}
\end{equation}  
In the above equation the integration is taken from the classical inner turning point $r_{\rm in}$ to infinity. This phase integral is closely related to the number of bound states supported by the interaction potential  \cite{Gribakin:1993}:
\begin{equation}
	N= \lfloor\Phi/\pi+\frac{3}{8}\rfloor
\end{equation}
There is also a one-to-one correspondence between the scattering length and the energy of the last bound state below the dissociation limit \cite{Gao:2001,Gao:2004,Kitagawa:2008}:
in the simplest case \cite{Blatt:1949} of very large $a$, $E_{-1}=-\hbar^2/{2\mu a^2}$.

For systems with large reduced masses and deep potentials, such as RbYb, the scattering
length is very sensitive to the variation of  the BO interaction potential.
When we parametrize $\Phi $ with a uniform, dimensionless, scaling parameter $\lambda$, which scales either 
the reduced mass of the system {\em  or} the interaction potential, then one cycle of the scattering length, where $\Phi$ changes by $\pi$, takes approximately 
\begin{equation}
\frac{\Delta \lambda}{\lambda} \approx \frac{2}{N}
\end{equation}
for a large number of bound states $N$. In case of RbYb --- as we will show in  Section \ref{sec:results} --- a 3\% variation of the interaction potential changes the scattering length from $-\infty$ to $+\infty$ within one cycle. 
Similarly, a full cycle of scattering length variation occurs when the reduced mass is changed by $\Delta \mu \approx 2 \mu /N$.

Except for very light, few-electron dimers, such as He$_2$ the Born-Oppenheimer interaction potential  cannot be determined with current {\em ab initio} methods precisely enough to evaluate the scattering length reliably. In fact, the \emph{ab initio} methods for systems involving lanthanides are extremely demanding: the quantum chemistry treatment needs to use high quality methods to include dynamic electronic correlation- and relativistic effects.

A model potential with the correct number of bound states can be worked out thanks to extra information provided by the experiment. The most useful data can be  provided by two-color photoassociation which probes the bound states from the top of the interaction well. For a series of experimental bound-state energies for a single isotopomer we can obtain a series of potentials,
which correspond to states with a similar value of $\tan \Phi$ (and consequently the scattering length), but with the number of bound states differing by $\pm 1,\pm 2 \ldots$ from the real potential.
The crucial assumption needed to completely back out the information about the real potential and the  phase shift integral is that the product $\mu V(r)$ which appears in the Eq. \ref{Phi} be linear in mass and that $V(r)$ be mass independent. This assumption is fulfilled extremely well for diatomic molecules with large reduced masses, where mass-dependent corrections to the Born-Oppenheimer potentials are small. In such case one can assume that for {\em all} isotopic combinations the only dependence on reduced mass in Eq. (\ref{Phi}) is due to the $\sqrt{\mu}$ whereas the integral in Eq. (\ref{Phi}) is identical for all isotopic pairs, thus the $\Phi/\sqrt{\mu}$ ratio can be found. Hence, two-color photoassociation spectroscopy performed for two or more isotopic combinations can fix exactly the  $\Phi/\sqrt{\mu}$ ratio and, consequently, the number of bound states supported by the interaction potential and the positions of few highest bound states for all possible isotopic combinations.


\section{The model potential}

\begin{figure}
\includegraphics[width=0.5\textwidth]{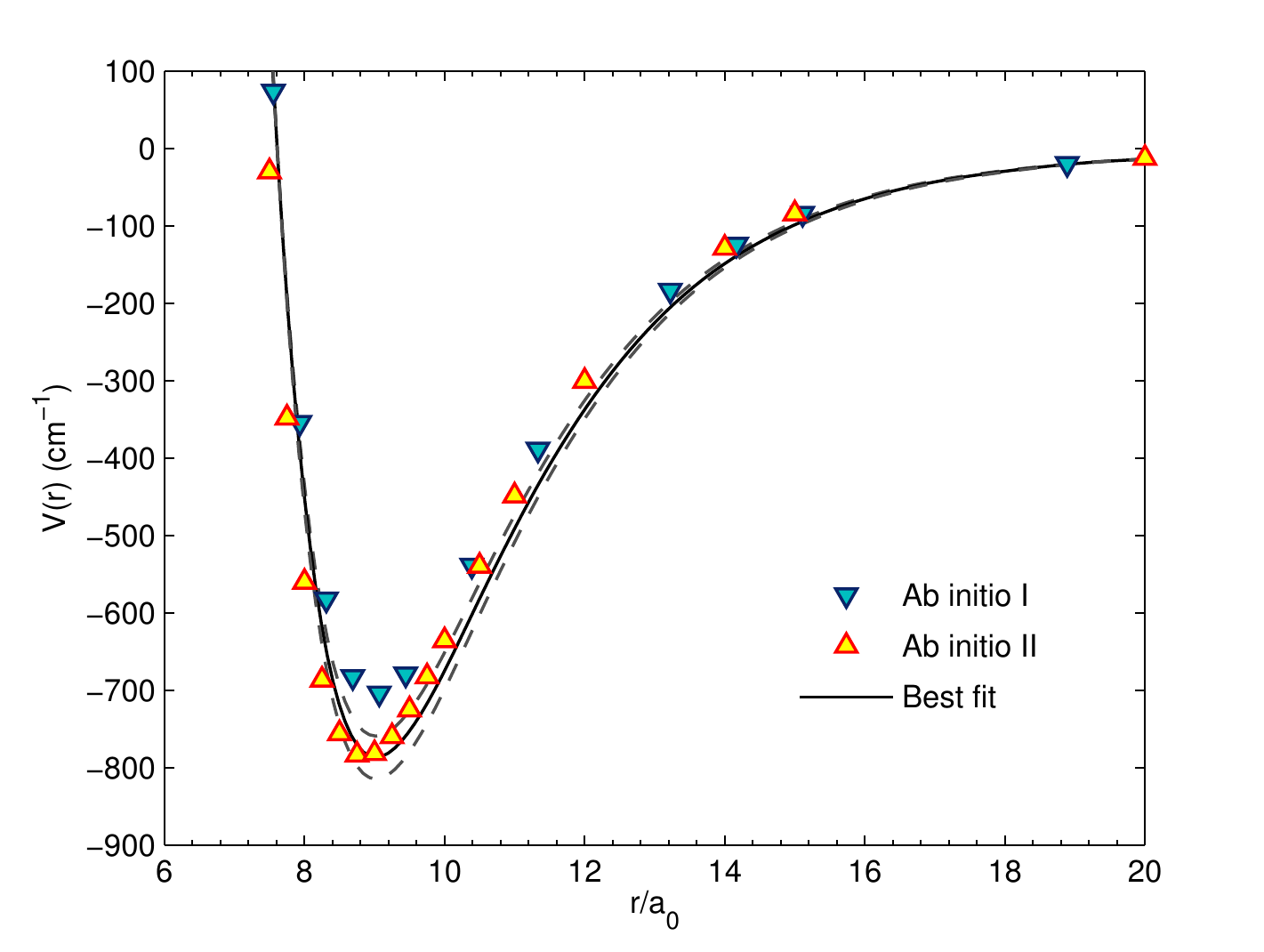}
\caption{(color online) Theoretical RbYb interaction potentials. Our two \emph{ab initio} potentials are shown as blue and orange triangles corresponding to the calculated potential points. `Ab initio I' is the base short range curve used in the potentials fitted to the experimental data, while `Ab initio II' is used as a benchmark of the obtained value of the potential depth $D_e$. The best potential obtained by fitting to the experimental data (see Section \ref{sec:results}) is shown as a black solid line. In terms of potential depth, it is in very good agreement with our benchmark potential `Ab initio II'. Additionally, we show two other fitted potentials supporting one fewer and one more vibrational bound states than the best fit potential as dashed lines. We use the other two potentials to evaluate possible systematic errors in our analysis.\label{fig:potentials}}
\end{figure}

Apart from fixing the real number of vibrational states we also provide a potential 
physically valid both at short and long ranges of interaction. To this end we introduce a model  
based on quantum chemistry {\em ab initio} calculation at short range $V_{\rm sh}(r)$
connected smoothly to the long-range analytical form of $-C_6r^{-6}-C_8r^{-8}$:
\begin{equation}
V(r)=dV_{\rm sh}(r)\left(1-f(r)\right) - f(r) \left( C_6r^{-6} + C_8r^{-8}\right)
\end{equation} using the switching function $f(r)$
introduced by Janssen {\em et al.} \cite{Janssen:2009}:
\begin{equation}
	f(r) = \left \{ 
	\begin{array}{l l}
	0 & r \le a \\
	\frac{1}{2}+\frac{1}{4}\sin (\pi x/2) \left( 3-\sin^2 (\pi x/2) \right) & a<r<b\\
	1 & b \le r \\
	\end{array}
	\right.
\end{equation}
with $x = \left( (r-a) + (r-b)\right)/(b-a)$. The parameters $a$ and $b$ define the switching range.

 The original short-range {\em ab initio} potential can be scaled by a factor $d$ in order to set the potential depth $D_e$. We vary $d$ and the $C_6$, $C_8$ coefficients to minimize the least-squares fitting error between the calculated and measured photoassociation line positions.
The short range part is sampled on a fine grid and interpolated using cubic splines to calculate the values of the potential at arbitrary points. The switching distance has been optimized by hand \footnote{The automated fitting procedure tended to make the transition either unphysically steep or even reverse the transition bounds which caused the program to crash.} and ranges from $a=17$ a.u. to $b=20$ a.u. 
Below we give details of tests and calculations of the short-range interaction potential as well as the discussion of its long-range part.


\subsection{{\em Ab initio} calculations }

The {\em ab initio} calculations for molecules  such as Rb and Yb are very demanding for several reasons.
First, since there are many electrons moving around the very heavy nuclei
one can predict that the ``dynamic'' electronic correlation effects will be very large, therefore we have to employ the most accurate affordable quantum-chemistry ab-initio methods known. Coupled-cluster theory including singly-
and doubly excited states with noniterative inclusion of triply excited states, abbreviated as CCSD(T), is therefore the most reasonable choice. Secondly, a common problem with species containing  $ns^2$ atoms, such as alkaline earth atoms, is a strong contribution of nondynamic electronic correlation, related to the $6s-6p$ orbital mixing. Thus, the reliability of coupled-cluster calculations needs to be carefully monitored using the so-called {\sc T1} diagnostic test of Lee and Taylor \cite{LEE:1989}. Another  difficulty is the lack of
a correlation-consistent family of gaussian basis sets for both Yb and Rb atoms, in order to systematically follow the
error of the interaction energy caused by incompleteness of the basis-set expansion of the molecular wavefunction. For the same reason predicting the interaction energies at the limit of a complete basis set is still impossible for this system.   
Finally, it is also essential to include in such calculations the relativistic effects, due to the extremely large charge of the Yb nucleus.

In a previous study of the RbYb system S{\o}rensen and coworkers\cite{Sorensen:2009} have used a full 4-component relativistic coupled-cluster method to extensively study the ground state of this system with aug-cc-pVTZ type  basis set of Dyall \cite{Dyall:2009,Gomes}. The difference between their counterpoise-corrected well depth  ($D_e=749~{\rm cm}^{-1}$) and the non-counterpoise corrected one ($D_e=870~{\rm cm}^{-1}$) suggests a strongly imbalanced basis set. The error due to basis set incompleteness might be quite large - if the difference between CP-CCSD(T) and CCSD(T) were taken as  an error bound, the uncertainty of the number of bound states could be as large as 5-6. Thus, further studies of {\em ab initio} potentials are essential to 
reduce this discrepancy.

In this work we have used two independent approaches to calculate the {\em ab-initio} interaction potentials at short range. In the first approach we have employed the calculations based on the ECP28MDF pseudopotential for the Rb atom \cite{Lim:2005,Lim:2007} and the relativistic medium-core (60 electron frozen, 10 active) pseudopotential of Wang and Dolg \cite{Wang:1998:Yb}for Yb. We have modified the original basis set provided with the ECP28MDF pseudopotential by taking out completely the contraction coefficients
 extending the basis set by adding (manually optimized) functions with additional exponents $f=0.06186$, $g=1.36$ and $h=1.12$,
and adding additional  diffuse $s-h$ functions using the even-tempered scheme implemented in {\sc molpro}.  
In a similar manner we have modified  the original basis set for the Yb atom:  we have uncontracted the original functions, and added respectively  5, 3, 2   outermost $f,g,h$ exponents 
from the ANO-RCC basis set \cite{Roos:2008}, and  finally, diffuse $s-h$ even tempered functions.
To better account for dispersion effects which are crucial in RbYb (which is unbound without the dispersion)
we have added the bonding functions in the center of molecule ($spd=0.9, 0.3,0.1$; $fg=0.6,0.2$).
Using  such basis sets and ECPs  we have calculated counterpoise corrected interaction energies 
using the coupled-clusters method which includes singly- and doubly-excited determinants with noniterative triples correction
[CCSD(T)]. We have correlated 19 electrons (Yb: $5s^{2}5p^{6}6s^{2}$ Rb:
$4s^{2}4p^{6}5s^{1}$) in this calculation.
Since this approach is relatively inexpensive and we could afford to calculate many distances with a fine grid,
this method  was our ultimate choice for the equilibrium-range representation of the potential. 
The obtained potential curve is shown in Fig. \ref{fig:potentials}, labelled `Ab initio I' and this is the base short range \emph{ab initio} curve used to fit to the experimental data.

In our second approach  we employed high level methods of quantum chemistry combined with an all-electron basis set.
 Benefiting from the work of S{\o}rensen \cite{Sorensen:2009}, where the
CCSD(T) calculations with the Dirac hamiltonian were  performed, one can
notice that the correlation is the most important part in the description of
the RbYb ground state  and the relativistic effects can be saturated by
using so called spin-free Hamiltonians. Following their conclusions, our
main effort was  put in recovering the correlation. Therefore, the
CCSD(T) approach with 23 (Yb: $4f^{14}6s^{2}$, Rb: $4p^{6}5s^{1}$)
correlated electrons and the full virtual space was used. As for basis sets we have used  the ANO-RCC basis sets \cite{Roos:2008}, however, we have uncontracted them for both atoms.
Additionally, the mid-bond functions ($sp=0.9, 0.3,0.1$; $df=0.6,0.2$) were used to improve the description of the molecular wavefunction.
In order to avoid problems with basis set superposition error, the potential
was counterpoise-corrected.
 Finally, relativistic effects were included by  applying the DKH3 Hamiltonian \cite{Nakajima:2000}.
Calculations in an all-electron basis set approach were performed with the help of the NwChem package \cite{nwchem}.
The calculations in this second approach are much more expensive: the total basis set size was over 500 basis functions and we were able to investigate only the region near the bottom of the interaction well. Thus, they served only as the benchmark data with respect to $D_e$ extracted from experiment. 
The resulting potential curve is shown in Fig. \ref{fig:potentials}, labeled `Ab initio II'.


\subsection{Long range interactions}

\begin{table}[htbp]
  \centering
\begin{ruledtabular}
  \caption{A comparison of potential parameters: long range van der Waals coefficients $C_6$ and $C_8$, the potential depth $D_e$, the harmonic constant at equilibrium $\omega_e$ calculated for $^{87}$Rb$^{176}$Yb and the equilibrium distance $R_e$. The first three rows list RbYb ground state potential parameters reported previously in the literature: the \emph{ab initio} potential of Sorensen \emph{et al.} \cite{Sorensen:2009}, the recent long range coefficients from Porsev \cite{Porsev:2013} and those reported by Brue and Hutson \cite{Brue:2013} which were based on a previously published subset of the photoassociation data reported here. The values in the `Semi-empirical' row were calculated from atomic polarizabilities  (eqs. \ref{eq:c6} \& \ref{eq:c8}). `Ab initio I' and `Ab initio II' are values obtained directly from our two \emph{ab initio} calculations. Finally, `Experimental fit' row lists the values obtained from our best fit potentials.\label{tab:comparison} }
    \begin{tabular}{llllll}
          & $C_6$ & $C_8$  & $D_e$  & $\omega_e$ & $R_e$ \\
	& (a.u.) & ($\times 10^5$ a.u) & (cm$^{-1}$) & (cm$^{-1}$) & (a.u.) \\
	\hline
    Sorensen et al. \cite{Sorensen:2009} \footnote{Data taken from Table IV in \cite{Sorensen:2009} calculated with the counterpoise-corrected CCSD(T) with 23 correlated electrons.} & --    & --    & 749   & 27.95\footnote{Value rescaled to a reduced mass corresponding to the $^{87}$Rb$^{176}$Yb pair} & 8.93 \\
    Porsev et al. \cite{Porsev:2013} & 2837(57) & 3.20(7) & --    & --    & -- \\
    Brue and Hutson \cite{Brue:2013a} & 2874.7 & 7.57\footnote{Calculated from a $C_8/C_6$ ratio} & 719.1 & --    & 9.28 \\
    Semi-empirical & 2825.9 & 3.38  & --    & --    & -- \\
    Ab initio 1 & --    & --    & 704.46 & 26.58 & 9.02 \\
    Ab initio 2 & --    & --    & 785.80 & 28.81 & 8.85 \\
    Experimental fit & 2837(13) & 4.6(0.9) & 787(18) & 29.7(4) & 9.02 \\
    \end{tabular}%
\end{ruledtabular}
\end{table}%

The asymptotic form of the interaction between two  atoms takes the form
\begin{equation}
V_{\rm int}(r)=-C_6 r^{-6} - C_8 r^{-8} - \ldots .
\label{vint_as}
\end{equation}
It is usually difficult to obtain the Van der Waals constants $C_6$, $C_8$ directly from \emph{ab initio} calculations due to their large inaccuracy in the long range. They can, however, be related to monomer properties
via Casimir-Polder type integrals:
\begin{equation}
C_6 = \frac{3}{\pi} \int_0^\infty \alpha_{\rm A}(i u) \alpha_{\rm B}(i u) d u  \label{eq:c6}
\end{equation}
and
\begin{equation}
C_8 = \frac{15}{2\pi} \Big(\int_0^\infty  \alpha_{\rm A} (iu) \beta_{\rm B }(iu) du  + \int_0^\infty  \alpha_{\rm B} (iu) \beta_{\rm A }(iu) du  \Big). \label{eq:c8}
\end{equation}
In  above equations  $\alpha$ denotes the dipolar polarizability, while $\beta$ is the quadrupole polarizability. 
In order to evaluate these  integrals  we need both quantities evaluated at imaginary frequencies. The calculations 
of these quantities at present is routine for closed-shell systems with the so-called time-independent coupled-cluster (TI-CC)
response function \cite{Moszynski:2005,Korona:2006} implemented efficiently in the  {\sc molpro} \cite{molpro} package.
Using this code and  the ECP60 basis set described in the previous section we have evaluated 
$\alpha(i u)$ and $\beta(i u)$ for the ytterbium atom at 50 frequencies corresponding to Gauss-Legendre quadratures given by Derevianko {\it et al.}
\cite{Derevianko:2010}. The TI-CC response function in the static limit gives dipolar polarizability which is at the top of the error bound  limit of best 
estimates of Dzuba and Derevianko~\cite{Dzuba:2010}. The $C_6$ value for the Yb dimer calculated with TI-CC dynamic polarizabilities gives 2165 a.u.
which is 12.2\% more than the value derived from experiment by Borkowski {\it et al.}. In order to fix this inaccuracy we divided the dipolar polarizability $\alpha_{\rm Yb}(i u)$ by $\sqrt{2165/1929}\approx 1.059$. 
For the Rb atom we have used the dynamic polarizabilities given by Derevianko {\it et al.}  \cite{Derevianko:2010}. The value of $C_6$ for the Rb-Yb interaction obtained with these functions was found to be 2826 a.u. This value has been used in our potential as a starting value for the least-squares fitting procedure. The $C_8$ coefficients can be evaluated with smaller confidence: the quadrupole dipole polarizability of Rb atom has been taken from Ref. \cite{Skomorowski:2011} where it has been constructed from pseudostates obtained from multireference configuration interaction calculations. This quadrupole polarizability gives the static limit of 6876 a.u. which overestimates the reference value of Mitroy and Bromley \cite{Mitroy:2003} by 6.1\%. We have rescaled the Rb quadrupole polarizabilities to obtain the correct static limit and used them with Yb (corrected) dipole and quadrupole response functions to calculate $C_8=3.38\times10^5$ a.u. for Rb-Yb interaction. A comparison between the calculated parameters and ones obtained previously in literature is shown in Table \ref{tab:comparison}.

\section{Data analysis}
\label{sec:results}

The photoassociation data obtained from the experiment enable us to produce a theoretical potential curve that we could later use to calculate the interspecies scattering lengths. As discussed earlier, this requires both that the long range part of the potential is correct and also that the phase integral, or the number of states be correct, so that mass scaling is well satisfied. In our data analysis we have satisfied both of these conditions, first by fitting a series of potentials which all have the correct long range interaction, but support different numbers of bound states, and then by selecting the potential that best describes the different isotopes.

\subsection{Least-squares fitting procedure}

The best-fit potentials were obtained using the least-squares method, which relies on minimizing the quantity

\begin{equation}
	\chi^2 = \sum_{i=1}^{n} \frac{(\Delta_{{\rm th},i}-\Delta_{{\rm exp},i})^2}{u(\Delta_{{\rm exp},i})^2}
\end{equation}
which is a measure of the discrepancies between the theoretical ($\Delta_{{\rm th},i}$) and experimental ($\Delta_{{\rm exp},i}$) line positions with respect to the experimental uncertainties $u(\Delta_{{\rm exp},i})$. The theoretical energy levels are produced by numerically solving the radial Schr\"odinger equation 
\begin{widetext}
\begin{equation}
	 -\frac{\hbar^2}{2\mu_i}\frac{d^2}{dr^2}\Psi(r) + V(C_6, C_8, D_e; r)\Psi(r) + V_{\rm HF}(F_i; r) 
	+ \frac{\hbar^2}{2\mu_i r^2} R_i(R_i+1)  \Psi(r)  =  E_{{\rm th},i} \Psi(r) \label{eq:se}
\end{equation}
\end{widetext}
\noindent for the trial potential $V(C_6, C_8, D_e; r)$, rotational quantum number $R_i$ and angular momentum $F_i$.  The eigenvalue closest to the corresponding experimental value is chosen for comparison, and as long as the trial potential is good enough, this provides the correct eigenvalue selection. The line position is then calculated via $\Delta_{{\rm th},i} = E_{{\rm th},i}/h - kT\left(R+1/2\right)/h$ \cite{Jones:2000}, where $T$ is the fitted sample temperature and $k$ and $h$ are the Botzmann and Planck constants respectively. The three parameters of the potential --- $C_6$, $C_8$ and potential depth --- are optimized using the well known Marquardt-Levenberg fitting algorithm along with the sample temperature $T$. Here we assume that the temperature of the atoms has not varied significantly for all measured two-photon resonances. This assumption is reasonable since all experimental measurements were performed under similar conditions. In addition, the temperature of Yb is expected to be independent of the isotope since the four bosonic isotopes used in the context of the work reported here are all lacking hyperfine structure. The limitation of the Levenberg-Marquardt fitting method is that it requires that the initial potential already reproduces the experimental energy levels closely enough that the correct eigenvalues can be selected as the `theoretical' values.

\begin{figure}
	\includegraphics[width=0.5\textwidth]{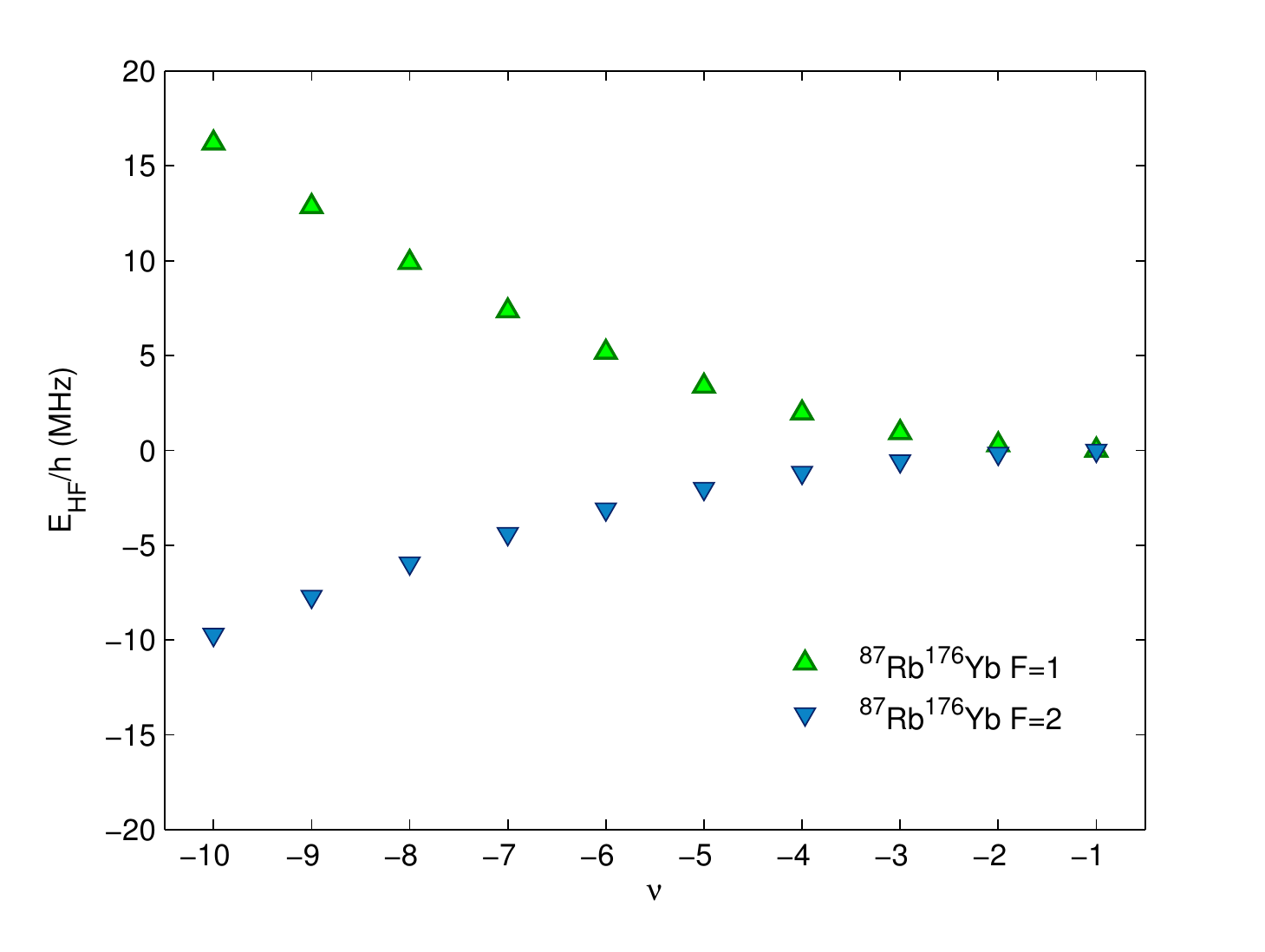}
	\caption{(color online) The impact of the hyperfine interaction on the bound state energies calculated for eight of the highest lying states of the ground state $^{87}$Rb$^{176}$Yb molecule. The molecular hyperfine shift $E_{\rm HF}$ was calculated by solving the radial Schr\"odinger equation (eq. \ref{eq:se}) with and without the hyperfine terms. \label{fig:hf}}
\end{figure}

Finally, the interaction potential is complemented by a hyperfine interaction potential $V_{\rm HF}(F;r)$, which takes into account the $r$-dependent change in the hyperfine constant in the RbYb molecule. This potential is explicitly dependent on the $F$-state of the molecule (bearing in mind that the Rb atom is in the $^{2}$S${_{1/2}}$ and the Yb atom is in the $^1$S$_0$ state):
\begin{equation}
	V_{\rm HF} = s \frac{1}{2} \left( F(F+1) - I(I+1) - S(S+1) \right) \Delta \zeta \,,
\end{equation}
where $\Delta \zeta$ is a molecular hyperfine correction function calculated analogously to ref.\cite{Zuchowski:2010:RbSr}, $s$ is an isotope dependent scaling factor proportional to the atomic hyperine splitting. The electron spin $S=1/2$, while nuclear spin $I=5/2$ for $^{85}$Rb and $I=3/2$ for $^{87}$Rb giving total atomic angular momentum $F=2,3$ or $F=1,2$ respectively.

We find that the impact of the hyperfine corrections on the energy levels is small and on the order of few MHz, as shown in Fig.~\ref{fig:hf}. As such, it was not critical for the quality of the fit --- the inclusion of these effects causes the $\chi^2$ to change by no more than 0.2. On the other hand, the inclusion of the temperature shift into the fitting reduced the $\chi^2$ factor of our best fit potential by over 35\%.

\begin{table}
\caption{A list of ten fits of the potential parameters to the experimentally determined line positions. Each of these fits supports a different number of bound states (ranging from $N=61$ to $N=70$). The potential with the lowest $\chi^2$ -- $N=66$ -- is the one that mass-scales best. We use the neighboring potentials ($N=65$ and $N=67$)  as a measure of the systematic (or model-dependent) error on the reported potential parameters.\label{tab:fitlist}}
\begin{ruledtabular}
    \begin{tabular}{cccccccccc}
    $D_e$    & $S(D_e)$ & $C_6$    & $S(C_6)$ & $C_8$    & $S(C_8)$ & $T$     & $S(T)$  & $\chi^2$ & $N$ \\
\multicolumn{2}{c}{(cm$^{-1}$)}  &\multicolumn{2}{c}{(a.u.)}  &\multicolumn{2}{c}{($\times 10^5$ a.u.)}  &\multicolumn{2}{c}{($\mu K$)} & \\
\hline
    656.0 & 6.2   & -2746.1 & 8.2   & -3.80 & 0.82  & 172   & 162   & 38.1  & 61 \\
    681.1 & 6.3   & -2765.3 & 8.2   & -3.98 & 0.84  & 196   & 162   & 25.6  & 62 \\
    706.8 & 6.5   & -2784.0 & 8.2   & -4.15 & 0.85  & 220   & 162   & 16.1  & 63 \\
    733.1 & 6.7   & -2802.2 & 8.3   & -4.32 & 0.87  & 244   & 162   & 9.6   & 64 \\
    759.9 & 6.8   & -2820.0 & 8.3   & -4.47 & 0.88  & 270   & 162   & 6.0   & 65 \\
    787.4 & 7.0   & -2837.2 & 8.3   & -4.62 & 0.89  & 296   & 162   & 5.3   & 66 \\
    815.4 & 7.2   & -2854.0 & 8.4   & -4.75 & 0.90  & 323   & 162   & 7.5   & 67 \\
    844.2 & 7.3   & -2870.3 & 8.4   & -4.86 & 0.91  & 350   & 162   & 12.4  & 68 \\
    873.5 & 7.5   & -2886.2 & 8.4   & -4.97 & 0.92  & 377   & 162   & 20.1  & 69 \\
    903.6 & 7.6   & -2901.6 & 8.5   & -5.05 & 0.93  & 404   & 162   & 30.5  & 70 \\
    \end{tabular}%
\end{ruledtabular}
\end{table}%

\begin{table}
	\caption{ Error budget for the fitted potential parameters --- the depth $D_e$ and the two van der Waals coefficients $C_6$ and $C_8$. The uncertainty given is calculated using method B from the fit uncertainty (as the statistical error) and the systematic error, which is calculated from the difference between the parameters in the chosen best fit potential and parameters for potentials supporting one less, and one more vibrational state. See text.}
\begin{ruledtabular}
    \begin{tabular}{rrrrrrrr}
     & $p$ & $S(p)$ & $\Delta p$ & $u(p)$ & Recommended \\
	\hline
    $D_e$ (cm$^{-1}$)   & 787.36 &  7.00  & 28.07 & 17.66 & 787(18) \\
    $C_6$ (a.u.)   & 2837.19 & 8.33  & 17.24 & 12.98 & 2837(13) \\
    $C_8$ (10$^5$ a.u.)   & 4.62 & 0.89 & 0.15 & 0.90 & 4.6(0.9) \\
    \end{tabular}%
\end{ruledtabular}
	\label{tab:parameters}
\end{table}

\subsection{The number of states}

Table~\ref{tab:fitlist} lists the parameters of ten best-fit potentials, each supporting a different number of vibrational states or, equivalently, $\Phi$ differing by an integer $k \pi$. The parameters shown are their depth $D_e$, the van der Waals coefficients $C_6$ and $C_8$ and the fitted sample temperature $T$. The quality of the fit is determined by the $\chi^2$ factor. Finally, for each potential we give the number $N$ of bound states supported for the case of $^{87}$Rb$^{176}$Yb and R=0. Also, each of the fitted parameters is paired with its corresponding statistical error calculated from the fit.

The long range parameters $C_6$ and $C_8$ are to a great extent fixed by the level spacings alone (even data for one isotope would suffice). The main difference between the fitted potentials is their depths, which vary from 656.0~cm$^{-1}$ to 903.6~cm$^{-1}$. Most of our photoassociation data has been recorded for the $^{87}$Rb$^{176}$Yb pair, which fixes the long range part of the potential very well. The remaining 13 lines, recorded for $^{87}$Rb combined with $^{170}$Yb, $^{172}$Yb, and $^{174}$Yb enable us to choose the potential that best reflects the mass scaling behavior of the RbYb molecule. The best fit quality is obtained for the 787.4~cm$^{-1}$ potential that supports $N=66$ bound states (for $^{87}$Rb$^{176}$Yb and $R=0$).
The selected potential clearly has the lowest $\chi^2$. Since, however, it is only 13\% lower than that of the $N=65$ potential and 40\% lower than with the $N=67$ potential, we believe that the $N=66$ potential has to be treated with caution. We will therefore use the differences in parameters and scattering lengths calculated from the best fit and the two second best fit potentials as a measure of an additional \emph{systematic} error. Thus, the reported experimental uncertainties will reflect our uncertainty of the number of bound states as well. The value for the temperature obtained from the fit is consistent with the experimental conditions.

A comparison of experimental and theoretical bound state energies is shown in Table \ref{tab:data}. The model describes the experimental data well within the error bars in most cases, as evidenced by the very low $\chi^2$~=~5.3. The expected value in our case is 47, which suggests that the experimental error bars may have been estimated very conservatively.

\subsection{Error analysis}

For each parameter or scattering length, the systematic error was calculated by taking the value predicted by our chosen best fit model and comparing it to one predicted by best fit potentials supporting one less and one more bound state as discussed earlier. The difference (whichever larger) was then taken as a measure of the systematic error. The dependence of the potential parameters on the number of supported states $N$ is shown in Table~\ref{tab:fitlist}. The statistical error for the potential parameters was directly computed by the fitting procedure, along with their correlation matrix $\rho$. It should be noted that all three fitting parameters, ie. the potential depth $D_e$, $C_6$ and $C_8$ tend to shift the energy levels down when their magnitude is increased. This causes significant correlation between them and, in fact, the correlation matrix for the respective parameters ($D_e$, $C_6$, $C_8$) of the chosen best fit potential reads:
\begin{equation}
	\rho = \left( \begin{array}{c c c}
	1 & -0.282134 & -0.998663 \\
	-0.282134 & 1 & 0.232436 \\
	-0.998663 & 0.232436 & 1 \\
	\end{array}
\right)\,.
\end{equation}
The correlation between the potential depth and $C_8$ is especially striking. The explanation is that the energy level spacing for the measured states is mostly determined by the van der Waals interaction parameter $C_6$. While $C_8$ provides an important correction to this spacing, it also strongly influences the phase integral $\Phi$, as does the potential depth. The complete error budget for the potential parameters has been laid out in Table \ref{tab:parameters}.

To calculate the statistical uncertainty $S(a)$ of a scattering length $a$ (see next section), one also has to take into account those correlations. We use a standard expression for a combined statistical error applied to a scattering length $a$ posed as a function of fitted parameters $D_e$, $C_6$, and $C_8$: 
\begin{equation}
	S^2(a) = \sum_{i,j} \frac{\partial a}{\partial p_i}\frac{\partial a}{\partial p_j} \rho_{i,j} S_i S_j \, ,
\end{equation}
where $p_i$ denote the potential parameters.

The derivatives in the above expressions were evaluated numerically. It turns out that the inclusion of the three covariance terms is necessary for an accurate determination of the statistical uncertainties of the scattering lengths -- otherwise they are overestimated by about an order of magnitude. The final combined uncertainty $u(a)$ is evaluated as usual: 
$u^2(a) = S^2(a) + \Delta a ^2/3$, where $\Delta a$ is the systematic error discussed earlier.

\section{Results and discussion}

\begin{table*}[htbp]
  \centering
  \caption{ Scattering lengths in the RbYb system. For each isotopic pair we give the value of the scattering length calculated by solving the Schr\"odinger equation (column `$a$') without the hyperfine term in order to provide a hyperfine independent value. We also show the error budget for each reported value by giving the statistical error $S(a)$, the systematic error $\Delta a$ and the resulting standard `Method B' uncertainty $u(a)$. The resulting data is in agreement with the qualitative experimental data \cite{Baumer:2011, Tassy:2010}.}
\begin{ruledtabular}
    \begin{tabular}{rrrrrrrrrr}
    Rb    & Yb    & $a$     & $S(a)$ & $\Delta a$ & $u(a)$ & Exp.  & Recommended & \multicolumn{2}{c}{$\Delta a_{\rm HF}(F)$} \\
          &       &       &       &       &       &       &       & F=2   & F=3 \\
\hline
    85    & 168   & 229.80 & 5.13  & 18.08 & 11.63 &       & +230(12) & -0.21 & 0.15 \\
    85    & 170   & 139.87 & 1.67  & 4.89  & 3.28  &       & +139.9(3.3) & -0.07 & 0.05 \\
    85    & 171   & 117.33 & 1.30  & 3.34  & 2.32  &       & +117.3(2.3) & -0.05 & 0.04 \\
    85    & 172   & 99.65 & 1.07  & 2.52  & 1.81  &       & +99.7(1.9) & -0.04 & 0.03 \\
    85    & 173   & 84.30 & 0.97  & 2.05  & 1.53  &       & +84.3(1.6) & -0.04 & 0.03 \\
    85    & 174   & 69.86 & 0.97  & 1.78  & 1.41  &       & +69.9(1.5) & -0.04 & 0.03 \\
    85    & 176   & 38.98 & 1.23  & 1.58  & 1.53  &       & +39.0(1.6) & -0.05 & 0.03 \\
\hline
          &       &       &       &       &       &       &       & F=1   & F=2 \\
\hline
    87    & 168   & 39.19 & 1.21  & 1.58  & 1.51  &       & +39.2(1.6) & -0.12 & 0.07 \\
    87    & 170   & -11.45 & 2.31  & 1.78  & 2.52  & 0     & -11.5(2.5) & -0.22 & 0.13 \\
    87    & 171   & -58.90 & 4.17  & 2.16  & 4.35  &       & -58.9(4.4) & -0.40 & 0.24 \\
    87    & 172   & -160.70 & 10.84 & 3.07  & 10.98 &       & -161(11) & -1.05 & 0.62 \\
    87    & 173   & -625.67 & 87.41 & 5.88  & 87.48 &       & -626(88) & -8.58 & 5.05 \\
    87    & 174   & 880.26 & 114.33 & 17.53 & 114.77 & +500  & +880(120) & -11.02 & 6.76 \\
    87    & 176   & 216.85 & 4.41  & 2.49  & 4.64  &       & +216.8(4.7) & -0.44 & 0.26 \\
    \end{tabular}%

\end{ruledtabular}
  \label{tab:scatlens}%
\end{table*}%

\begin{figure}
\includegraphics[width=0.5\textwidth]{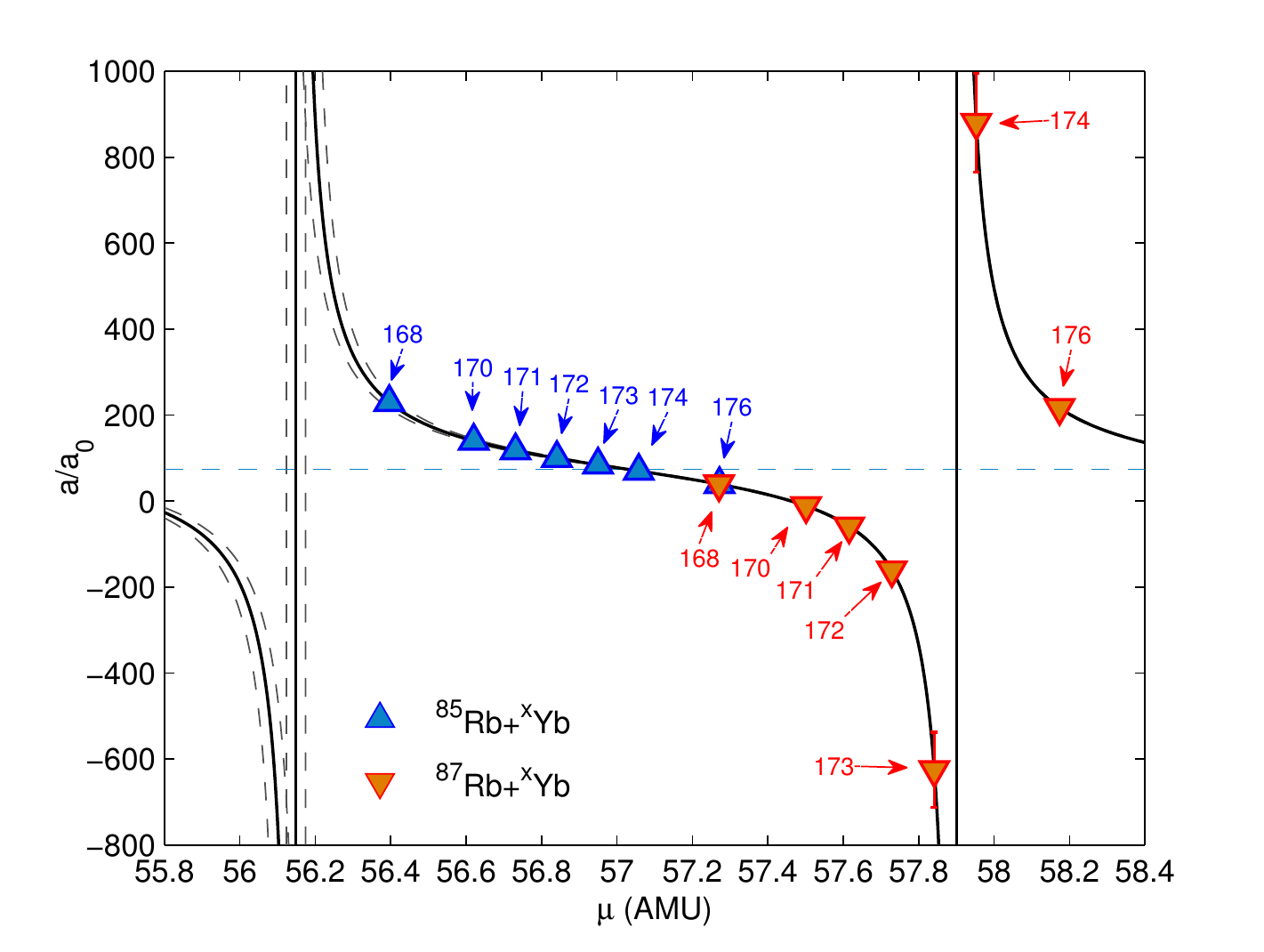}
\caption{(Color online) Mass dependence of the scattering length. The black line is the scattering length numerically calculated as a function of the reduced mass of the RbYb molecule. In the theoretical treatmenr introduced by Gribakin and Flambaum this dependence is given by the tangent function \cite{Gribakin:1993} (see eq.~\ref{eq:gribakin}) shifted by a quantity called the mean scattering length $\bar a$ (blue dashed line).
The  dashed black lines are similar predictions based on potentials supporting one less and one more vibrational state (in $^{87}$Rb$^{176}$Yb). \label{fig:tangent}}
\end{figure}

The values of the scattering lengths along with their respective error budgets are given in Table \ref{tab:scatlens}. These were evaluated by directly solving the same Schr\"odinger equation (eq.~\ref{eq:se}) as used during fitting, except for the hyperfine term. This way, a spin-independent value is obtained.
 The range of values covered by various $^{87}$Rb-$^{x}$Yb combinations includes both large magnitudes, as it is in the case of the large negative scattering length of the $^{87}$Rb-$^{173}$Yb pair, as well as the highly repulsive case of $^{87}$Rb and $^{174}$Yb, where the interaction is so large, that it can cause severe distortion of the atomic cloud when both species are trapped together, as confirmed in experiment \cite{Baumer:2011}. On the other hand $^{87}$Rb and $^{170}$Yb are characterised by a very small negative scattering length, which explains their extremely inefficient sympathetic cooling.

A different situation is seen in $^{85}$Rb-$^{x}$Yb combinations, where the scattering lengths are all moderate and positive -- they only range from 36.7~a.u. to 212~a.u. and are all situated around the mean scattering length $\bar a$ (see fig. \ref{fig:tangent}). It is worth noting that when the scattering length is close to $\bar a$, as it is the case in $^{85}$Rb-$^{173}$Yb and $^{85}$Rb-$^{174}$Yb, then according to Gao's theory \cite{Gao:2001}, there is a $d$-wave shape resonance close to zero scattering energy.

For completeness, we have also calculated the hyperfine corrections $\Delta a_{\rm HF}(F)$ to the scattering lengths. They are also listed in Table~\ref{tab:scatlens} for the possible values of the total spin $F$. In our case these corrections are all significantly smaller than the error bars of the scattering lengths themselves and they are much larger in $^{87}$Rb-$^x$Yb pairs than $^{85}$Rb-$^x$Yb because of the smaller hyperfine splitting in the latter systems. The spin-dependent scattering length can be evaluated via $a(F) = a + \Delta a_{\rm HF}(F)$.

We have also calculated the thermalization rates for the RbYb system in order to compare the predictions of the model with experiment. The elastic scattering cross sections for a given collision energy $\varepsilon$ are 
\begin{equation}
	\sigma(\varepsilon) = \frac{4\pi \hbar^2}{2\mu \varepsilon} \sum_{R=0}^{\infty} (2R+1) \sin^2(\eta_R) 
\end{equation}
where $\eta_R$ is the scattering wavefunction phase for a the partial wave $R$. The calculation of the scattering cross sections has been carried out using the {\sc molscat} \cite{molscat:v15} package. To obtain the thermalization rates in a thermal cloud we have calculated the thermal average of the cross sections using the Boltzmann distribution
\begin{equation}
	\left \langle \sigma \right \rangle = \frac{1}{k_B T} \int_0^{\infty} \sigma(\varepsilon) \exp(-\varepsilon/k_B T) d \varepsilon \, .
\end{equation}
The experimental data on the thermalization rates \cite{Tassy:2010} was only given relative to the thermalization rate  in the $^{87}$Rb$^{176}$Yb pair. Since the thermalization rate is directly proportional to the elastic scattering cross section, we can compare the ratio $\left \langle \sigma \right \rangle/\left \langle \sigma_{176} \right \rangle$ to the experimental data. In the experiment the temperature was estimated to be T=50~$\mu$K. We have found that the theoretical predictions agree with experimental data qualitatively -- the calculations confirm that the scattering cross section between $^{87}$Rb and $^{170}$Yb is very small, about three orders of magnitude lower than the $^{87}$Rb$^{176}$Yb pair, as evidenced by their extremely low thermalization rate. On the other hand, the thermalized scattering cross section between  $^{87}$Rb and $^{174}$Yb is two to three times larger and in experiment, the large interaction causes distortion of the atomic cloud \cite{Baumer:2011}. 


\section{ Conclusion}
In this paper we have determined an accurate model potential of the ground state of RbYb dimer
and the scattering lengths for all possible isotopic combinations. To this end we have 
introduced a potential which at short range uses \emph{ab initio} data which are smoothly 
connected to the analytic long-range form  $-C_6r^{-6}-C_8r^{-8}$. The 
short-range potential has been parametrized with a uniform scaling parameter $d$ which rescales the \emph{ab initio} data. With data provided from two-color photoassociation spectroscopy for $^{87}$Rb$^{176}$Yb, $^{87}$Rb$^{174}$Yb,  $^{87}$Rb$^{172}$Yb
and  $^{87}$Rb$^{170}$Yb isotopic mixtures (for $R=0,1$ rotational states), we have optimised the $C_6$, $C_8$ and $d$
parameters of the model  potential to minimise the least square error between predicted top-bound states  and the experimental values. In the calculation  of line positions we have included both the temperature and molecular hyperfine effects.
The recommended potential has the well depth $D_e = 787.4$ cm$^{-1}$ in very good agreement with our state-of-the-art {\em ab initio} 
calculation based on the Douglass-Kroll-Hess approximation (786 cm$^{-1}$) and previous, fully relativistic calculations of S{\o}rensen and coworkers \cite{Sorensen:2009}.

The scattering lengths in all  RbYb systems span a broad range of values. For two of isotopic combinations, $^{87}$Rb$^{174}$Yb and  $^{87}$Rb$^{173}$Yb, are nearly at the pole of scattering length: the first is  large and positive, the latter -- large and negative.
The previous experimental value of the scattering length of $^{87}$Rb$^{170}$Yb has also been confirmed 
by these studies. On the other hand the isotopic combinations involving $^{85}$Rb all have moderate scattering lengths close to the mean scattering length for this system. This suggests that $^{85}$Rb is a good candidate for the production of stable binary Bose-Einstein condensates (or quantum degenerate Bose-Fermi mixtures) with all ytterbium  mind that 85Rb Bose-Einstein condensates are only stable close to a homonuclear Feshbach resonance \cite{Cornish:2000}. For the system with lowest reduced mass $^{85}$Rb$^{168}$Yb the scattering length is still moderately far from the pole.

Two systems, $^{85}$Rb$^{173}$Yb and $^{85}$Rb$^{174}$Yb have their scattering length very close to $\bar{a}$ which is a condition for the shape resonance
in the $d$ partial wave. This feature can be exploited in future experiments: for example for these specific isotopic mixtures one could expect shape-resonance enhanced photoassociation from the rotationaly excited state ($R=2$) of the RbYb molecule. The model potential derived here is important for exploring the possibilities of manipulation of collisional properties
of the RbYb system: we will use it in future investigations of magnetically and optically tunable Feshbach resonances in the RbYb system.

\acknowledgments

This work has been partially supported by the Foundation for Polish Science TEAM Project \emph{Precise Optical Control and Metrology of Quantum Systems} and
 the Foundation for Polish Science  Homing Plus Programme no. 2011 - 3/14 cofinanced by the European Regional Development Fund. The project is a part of an ongoing research program of the National Laboratory FAMO in Toru\'n, Poland. The authors also acknowledge the Wroclaw Networking and Supercomputing Center, Academic Computer Center Cyfronet AGH and SURFsara (Amsterdam) for generous allotment of computer time. 
PSJ acknowledges support from an AFOSR MURI Grant No. FA9550-09-1-0617. This research was  supported in part by the National Science Foundation under Grant No. PHY11-25915.

\bibliography{my_refs}

\end{document}